\documentclass[twocolumn]{aastex6}

\begin{document}


\title{The extended Northern ROSAT Galaxy Cluster
   Survey (NORAS II)\\ I. Survey Construction and First Results}


\author{Hans B\"ohringer\altaffilmark{1}, Gayoung Chon\altaffilmark{1}, 
           J\"org Retzlaff\altaffilmark{2}, Joachim Tr\"umper\altaffilmark{1}, 
           Klaus Meisenheimer\altaffilmark{3}\, and Norbert Schartel\altaffilmark{4}}
\affil{}


\altaffiltext{1}{Max-Planck-Institut f\"ur extraterrestrische Physik, 
                 D-85748 Garching, Germany}
\altaffiltext{2}{ESO, D-85748 Garching, Germany}
\altaffiltext{3}{Max-Planck-Institut f\"ur Astronomy, K\"onigstuhl 17, 
                 D-69117 Heidelberg, Germany}
\altaffiltext{4}{ESAC, Camino Bajo del Castillo,
                   Villanueva de la Ca\~nada, 28692 Madrid, Spain}



\begin{abstract}
As the largest, clearly defined building blocks of our Universe, galaxy clusters are 
interesting astrophysical laboratories and important probes for cosmology. 
X-ray surveys for galaxy clusters provide one of the best ways to characterise 
the population of galaxy clusters. We provide a description
of the construction of the {\sf NORAS II} galaxy cluster survey based on
X-ray data from the northern part of the ROSAT All-Sky Survey. 
{\sf NORAS II} extends the {\sf NORAS} survey down to a flux limit 
of $1.8 \times 10^{-12}$ erg s$^{-1}$ cm$^{-2}$ (0.1 - 2.4 keV) increasing 
the sample size by about a factor of two. The {\sf NORAS II} cluster survey now
reaches the same quality and depth of its counterpart, 
the Southern {\sf REFLEX II} survey, allowing
us to combine the two complementary surveys.  
The paper provides information on the determination of the 
cluster X-ray parameters, the identification process of the X-ray 
sources, the statistics of the survey, and the construction of the survey 
selection function, which we provide in numerical format. 
Currently {\sf NORAS II} contains 860 clusters 
with a median redshift of $z = 0.102$.  
We provide a number of statistical functions 
including the logN-logS and the X-ray luminosity function and compare these
to the results from the complementary {\sf REFLEX II} survey. Using the  
{\sf NORAS II} sample to constrain the cosmological parameters, $\sigma_8$ and
$\Omega_m$, yields results perfectly consistent with those of {\sf REFLEX II}.
Overall, the results show that the two hemisphere samples, {\sf NORAS II} and 
{\sf REFLEX II}, can be combined without problems to an all-sky sample, just
excluding the Zone-of-Avoidance.
\end{abstract}




\section{Introduction} \label{sec:intro}

To explore the structure and evolution of our Universe we
have to rely on visible objects and observable physical processes.
Some of the most interesting systems for such studies are
galaxy clusters, the largest properly defined astronomical objects.
They allow us to trace the large scale matter distribution
(e.g. Peacock \& West 1992, Einsato et al. 1993, Schuecker et al.
2001, B\"ohringer et al. 2015) and to test cosmological models
(e.g. Perenod 1980, Borgani et al. 2001, Henry et al. 2004, 2009,
Vikhlinin et al. 2009, B\"ohringer et al. 2014,
Planck Collaboration 2014, 2016).
They are also interesting laboratories to study the galaxy 
population and the physics of the intergalactic medium 
(e.g. Mulchey, Dressler \& Oemler 2004, Voit 2005, 
Markevitch \& Vikhlinin 2007, Kravtsov \& Borgani 2012, Eckert et al. 2015,
Bykov et al. 2015, Churazov et al. 2016, Arevalo, et al. 2016).
Such studies require the systematic construction of statistically
complete, well defined samples of galaxy clusters.

The detection and characterisation of galaxy clusters by their
X-ray emission is still one of the most advanced techniques for
building large statistical galaxy cluster samples. Contrary
to the optical appearance of clusters, where they have to be
characterised as a collection of galaxies, X-rays show clusters as
single entities. The X-rays further provide evidence for a deep 
gravitational potential that binds the matter of the cluster. 
In addition X-ray surveys yield cluster samples which are approximately 
mass selected, since the X-ray luminosity is tightly correlated
with cluster mass (e.g. Pratt et al. 2009, B\"ohringer et al. 2012) 
and the X-ray emission is highly peaked at the center such that projection 
effects are minimized. Progress has recently been made 
with optical cluster surveys, in particular due to the much larger number 
statistics (e.g. Rozo et al. 2010, Rozo \& Rykoff 2014). We have also seen
major progress with large 
surveys in the millimeter regime using the Sunyaev-Zel'dovich effect 
(Reichardt et al. 2013, Hasselfield et al. 2013,
de Haan et al. 2016, PLANCK-Collaboration, 2016a, 2016b, Bleem et al. 2016).
The cluster detection at these wavelengths is also based on
the signature of the hot intracluster medium in clusters,
with a signal less affected by cluster cooling cores (e.g. Motl et al.
2005).

The {\sf NORAS} project is based on the {\sf ROSAT} All-Sky Survey 
(RASS, Tr\"umper 1993) which is the only full sky survey conducted with
an imaging X-ray telescope. In the past we have already used 
the RASS for the construction of the cluster catalogues of
the {\sf NORAS I} project. While {\sf NORAS I} was as a first step
focussed on the identification of galaxy clusters among the RASS
X-ray sources showing a significant extent, the complementary 
{\sf REFLEX I} sample  in the southern sky was
strictly constructed as a flux-limited cluster sample. A major extension
of the {\sf REFLEX I} sample, which roughly doubles the number of clusters,
{\sf REFLEX II} (B\"ohringer et al. 2013), was recently completed. It is
by far the largest, high-quality sample of X-ray selected galaxy
clusters. Already the quality of the {\sf REFLEX I} sample has enabled us 
to perform a number of cosmological studies. We demonstrated
that it can provide reliable measures of the large-scale structure
(Collins et al. 2000, Schuecker et
al. 2001), constraining cosmological parameters 
(Schuecker et al.  2003a, b; B\"ohringer 2011) which were
in good agreement with subsequently published
results from {\sf WMAP} (e.g. Komatsu et al. 2011). The improved quality and
statistics of {\sf REFLEX II} provided refined cosmological constraints
(B\"ohringer et al. 2014), limits on the mass of neutrinos
(B\"ohringer \& Chon 2015), and confirmed the statistical biasing concept
of galaxy clusters with a precise measurement of the power spectrum of their
spatial distribution as a function of their mass (Balaguera-Antolinez et al. 2011, 2012). 
The {\sf REFLEX} data have also been used for statistical studies
of cluster properties and scaling relations involving studies
of statistical subsamples of {\sf REFLEX}  (e.g. Ortiz-Gil et al., 2004,
Kerscher et al. 2001, B\"ohringer et al. 2007, Croston et al. 2008,
Pratt et al. 2009, 2010, Arnaud et al. 2010, B\"ohringer et al. 2010).
We also constructed a catalogue of superclusters from the latest version
of the REFLEX catalogue comprising 164 superclusters including close pairs
of clusters (Chon et al. 2013). We used the established statistics of
this super-cluster sample to arrive at a more physically motivated and 
quantitative definition of superclusters, by introducing a new 
class of objects which we call superstes-clusters 
(where superstes is the Latin word for survivor), which denotes those 
large-scale structures that will 
survive the cosmic expansion and collapse in the future
in a $\Lambda$CDM cosmology (Chon et al. 2015). {\sf REFLEX II}
also allowed us to explore the matter density distribution
in the local Universe in the southern sky (B\"ohringer et al. 2015).

 During the work on the {\sf REFLEX} and {\sf NORAS} projects,
earlier subsamples of clusters from these surveys were published in
Ebeling et al. (1996, 1997, 1998), De Grandi et al. (1999), 
Cruddace et al. (2002). Further cluster samples based on the RASS
comprise Ebeling et al. (2001, 2002).

In this paper we describe the extension of the {\sf NORAS} survey to bring
it to the same depth, completeness, and quality as {\sf REFLEX II}. 
The two combined surveys, {\sf REFLEX II} and {\sf NORAS II},
that we termed CLASSIX 
(Cosmic Large-Scale Structure in X-ray) survey (B\"ohringer et al. 2016), 
now describe a homogeneous all-sky cluster sample which
just leaves out the so-called zone-of-avoidance ($\pm 20$ degrees
around the Galactic plane). The
combined survey was recently used to obtain statistical constraints
on magnetic fields in galaxy clusters through a correlation of rotation 
measures of distant polarised radio sources with the location of
X-ray luminous galaxy clusters in the sky (B\"ohringer et al. 2016).

The {\sf NORAS II} survey now reaches
a flux limit of $1.8 \times 10^{-12}$ erg s$^{-1}$ cm$^{-2}$  in the 0.1 - 2.4 keV 
band. Redshifts have been obtained for all of the 860 clusters in the 
{\sf NORAS II} catalogue, except for 25 clusters for which
observing campaigns are scheduled. Thus with  3\%
missing redshifts we can already obtain a very good view of the 
properties of the {\sf NORAS II} cluster sample and obtain some first results, 
which we present in this paper.

The paper is organised as follows. In chapter 2 we describe the survey
properties and in section 3 the identification procedure and the 
properties of the sources. Section 4 explains
the derivation of the selection function and section
5 provides some statistics on the survey. In section 6 we
present the X-ray luminosity function and in section 7
we show first results on constraints of cosmological parameters
from the {\sf NORAS II} survey. Section 8 provides a summary and the
conclusions.

For the derivation of distance dependent parameters we use a geometrically flat
$\Lambda$-cosmological model with $\Omega_m = 0.3$ and $h_{70} = H_0/70$ km s$^{-1}$ 
Mpc$^{-1}$ = 1. All uncertainties without further specifications refer to 1$\sigma$
confidence limits. 

\section{Survey Properties} \label{sec:stwo}

The {\sf NORAS II} survey covers the sky region north of the equator outside
the band of the Milky Way ($|b_{II}| \ge 20$ deg.) We also excise a region
around the nearby Virgo cluster of galaxies which extends over several
degrees on the sky, where the detection of background clusters is hampered
by  bright X-ray emission. This region is bounded in right ascension
by $RA = 185 - 191.25^o$ and in declination by $DEC = 6 - 15^o$
(an area of $\sim 53$ deg$^2$).With this excision the survey area 
covers 4.18 steradian (13519 deg$^{2}$, a fraction of 32.7\% of the sky). 
{\sf NORAS II} is based on the 
RASS product RASS III (Voges et al. 1999) which was also used for 
{\sf REFLEX II}. The {\sf NORAS II} survey was
constructed in an identical way as {\sf REFLEX II} with a nominal flux limit of 
$1.8 \times 10^{-12}$ erg s$^{-1}$ cm$^{-2}$. The northern sky has on
average slightly larger exposures than the southern sky, due to the
more frequent shut down times in the particle belts of the South Atlantic 
Anomaly affecting the southern sky. 
Fig. 1 shows a comparison of the exposure time distribution for
{\sf NORAS II} in comparison to {\sf REFLEX II}. Since the 
construction of the {\sf NORAS I} survey was based on the detection of
clusters as extended X-ray sources in an early version of the RASS
data set and since the detection of a significant extent in published
RASS results was imperfect, {\sf NORAS I} was not statistically complete. 
Therefore {\sf NORAS II} is now the first statistically well 
described cluster sample for the northern ROSAT sky.

\begin{figure}[ht!]
\figurenum{1}
\plotone{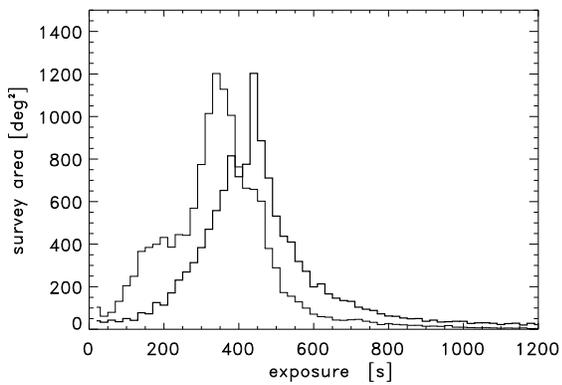}
\caption{Exposure time distribution of the {\sf NORAS II} survey
(thick line) in comparison to that of the {\sf REFLEX II} survey 
(thin line). Due to the smaller number of detector shut-off times
in the radiation belts in the Northern sky, the exposure coverage is
better than in the South.\label{fig1}}
\end{figure}

The source selection was based on an intermediate stage source
catalogue produced during the production of the RASS III data release
(Voges et al. 1999). 54,286 sources of this catalogue fall into the 
{\sf NORAS II} survey region. As in the case of {\sf REFLEX II}, we
reprocessed all these sources with the growth curve analysis (GCA) 
method, because we know that the standard analysis tuned for 
point sources tends to underestimate the flux of extended X-ray 
emission (B\"ohringer et al. 2000). A preliminary flux limit
of $1.4 \times 10^{-12}$ erg s$^{-1}$ cm$^{-2}$ was then imposed
on the reprocessed source catalogue yielding 3902 sources which
were further inspected in detail, as described below. The nominal
flux limit of $F_n \ge 1.8 \times 10^{-12}$ erg s$^{-1}$ cm$^{-2}$ has
been imposed after visual inspection, to allow sources to be recovered
in case the automatic flux measurement was not perfect.   

\section{X-ray parameters and source identification} \label{sec:sthree}

The X-ray fluxes and luminosities of the sources are based on
the source count rates determined with the growth curve analysis
method, which is described in detail in B\"ohringer et al. (2000, 2001).
The final values for flux and luminosity in a well defined aperture
are derived in three steps, which are described in detail in
B\"ohringer et al. (2013) and illustrated with a flow diagram
in their Figure 4. The three steps, which are briefly described
in the following, involve the determination of
(i) a preliminary, nominal flux without the knowledge of the
source redshift, (ii) a corrected flux taking the redshift information
into account, and (iii) a flux in a well defined aperture. 

\begin{itemize}

\item{{\bf nominal flux:} 
The count rate was measured with the GCA method
in the energy band defined by {\sf ROSAT} PSPC channels 
52 to 201 ($\sim 0.5$ to 2 keV), where the signal to noise 
is highest, and converted to a ``nominal'' flux, $F_n$, in the broad 
{\sf ROSAT} band, 0.1 to 2.4 keV. The aperture radius, $r_{out}$,  
for the count rate determination is defined by the radius
where the cumulative background-subtracted count rate
(growth curve) reaches a plateau without further significant
growth (for details see B\"ohringer et al. 2000, 2001).
The conversion from count rate to unabsorbed 
nominal X-ray flux, $F_n$, is performed by assuming a thermal 
plasma spectrum for a temperature of 5 keV, a metallicity of
0.3 of the solar value (Anders \& Grevesse 1989), a redshift of zero,
and an interstellar hydrogen column density given for the cluster
line-of-sight in the compilation by Dickey \& Lockman (1990).~\footnote{We 
have compared the results obtained with the interstellar
hydrogen column density compilation by Dickey \& Lockman (1990)
with those of the Bonn-Leiden-Argentine 21cm survey (Kalberla et al. 2005)
and found that the difference is of the order of a percent.
Because our survey has been constructed with a flux cut based
on the earlier results, we keep the older hydrogen column density
values for consistency of the cluster parameters with the selection
function.} The value of $F_n$ is used to make the flux cut independent of
any redshift information (since the redshift is not available
for all objects at the start of the survey).
}

\item{{\bf corrected flux and luminosity:} 
With known redshift, an improved flux calculation is performed
to obtain the unabsorbed X-ray flux, $F_x$. For this conversion
we use the cluster redshift and assume a 
temperature for the intracluster medium, that we derive
from the X-ray luminosity-temperature relation 

\begin{equation}
T = 3.31~ L_X^{0.332}~ h_{70}^{0.666} ~~,
\end{equation}

where $T$ is in keV and $L_X$ is measured
inside $r_{500}$~\footnote{$r_{500}$ is used for the
fiducial outer radius of the clusters, defined as the radius inside
which the mean mass density of the cluster is 500 times the critical
density of the Universe at the cluster redshift.}
in units of $10^{44}$ erg s$^{-1}$.
The $L_X$ - $T$ relation is taken from our {\sf REXCESS} study
(Pratt et al. 2009, B\"ohringer et al. 2012). Motivated by the results of
Reichert et al. (2011), we assume that the $L_X$ - $T$
relation has no redshift dependence. 
This correction is less than 5\% for sources with an
X-ray luminosity above $4 \times 10^{43}$ erg s$^{-1}$.
The use of a redshifted spectrum in the folding with the
instrument response is equivalent to a K-correction in
optical astronomy. Tables for the conversion from count rate
to flux and for the ``K-correction'' are given in B\"ohringer
et al. (2013) and are available on-line\footnote{They are available on 
CDS via anonymous ftp at cdsarc.u-strasbg.fr and on: 
http://www.mpe.mpe.de/$\sim$hxb/REFLEX}  
}  

\begin{figure}[ht!]
\figurenum{2}
\plotone{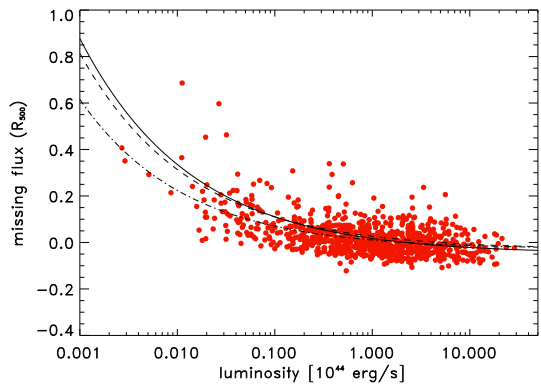}
\caption{Estimated missing flux as a function of X-ray luminosity for the
clusters in the {\sf NORAS II} sample. The effect of the ROSAT Survey 
point spread function has been included. The corrections are
largest for low luminosity objects. The solid line shows 
the best fitting power law function with linear offset (Eq. 2).
The dashed line shows the corresponding values for {\sf REFLEX II}.
The dotted-dashed line shows a fit, where 21 outliers have been
excised, to illustrate the large influence of the extreme outliers
on the fit.
\label{fig2}}
\end{figure}

\item{{\bf aperture flux and luminosity:} 
For general use we determine
the X-ray luminosity inside an aperture radius of $r_{500}$.
The procedure to convert the luminosities measured in
the aperture radius $r_{out}$ into luminosities inside
$r_{500}$ is described in detail in B\"ohringer et al. (2013).
The correction requires two pieces of information. $r_{500}$
is calculated from the cluster mass, which is in turn
estimated from the $L_X-M$ scaling relation (Eq. 5, see also 
B\"ohringer et al. 2014). In addition we have to correct for the effect
that, with the limited angular resolution of the survey,
flux originally inside $r_{500}$ is spread outside the
aperture. The half power radius of the RASS point spread 
function is almost 1.5 arcmin, and therefore this effect
cannot be neglected.

We use the {\sf NORAS II} cluster catalogue to illustrate the relevance
of this aperture flux correction in Fig.~\ref{fig2}. We find that
the missing flux depends most strongly on the cluster X-ray 
luminosity, as shown in the Figure.
The correction is most significant for the smaller, less luminous
systems. Due to their low surface brightness, we detect them with a
smaller extent. For clusters with luminosities above $10^{43}$ erg
s$^{-1}$, the mean correction is smaller than 2\%.
Fig.~\ref{fig2} also shows a functional fit of the 
average correction with the following form:

\begin{equation}
f_{miss} =  0.0677 \times L_{X,500}^{-0.379} -0.0504     ~~.
\end{equation} 

The result is almost identical to that obtained for the 
{\sf REFLEX II} survey as shown in Fig.~\ref{fig2}.
We have also fitted the data excluding the 21 most extreme outliers,
which resulted in a fit with a much lower reduced $\chi ^2$
(0.94 instead of 2.6) with the result 
$f_{miss} =  0.0387 \times L_{X,500}^{-0.407} -0.0282$. It is shown to 
illustrate the influence of a few extreme data points on the fit.
We are using the actual missing flux corrections, but not this fit
in our further analysis and therefore this problem does not affect
our results.}

\item{To model the cluster survey it is important to know 
the uncertainty in the flux and luminosity determination.
The main source of flux uncertainty is the Poisson statistics 
of the source photons. Thus it is not surprising that the uncertainty shows
the clearest dependency on the number of source photons detected.
We show this relation in  Fig.~\ref{fig3}, where we replaced the
photon number by the parameter of the product of flux and exposure. 
The function shows almost a square root dependence. It can be expressed
by: 

\begin{equation} 
f_{err} = 588.5 \times \left( F_X~ \times {\rm exposure} \right)^{-0.504}  ~~,
\end{equation}

where $F_X$ is given in units of $10^{-12}$ erg s$^{-1}$ cm$^{-2}$
and the exposure in sec.
This result is practically identical to the one obtained for
{\sf REFLEX II} as shown in Fig.~\ref{fig3}. 

The mean uncertainty of the flux is 17\%. Some data points 
that appear up-scattered from the average relation in Fig.~\ref{fig3} 
include those cases where sources have been deblended and where
the error increases due to the additional uncertainty in
deblending.}

\end{itemize}

\begin{figure}[h]
\figurenum{3}
\plotone{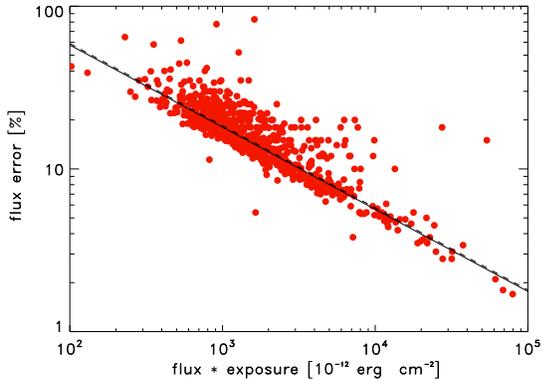}
\caption{Uncertainty of the flux determination for the
{\sf NORAS II} clusters as a function of the product
of flux and exposure time. This functional form is close to pure
Poisson statistics with a dependence on the square of the photon number.
The solid line shows the best power law fit
(Eq. 3). The dashed line shows the corresponding result for {\sf REFLEX II}.
\label{fig3}}
\end{figure}

We make use of two further parameters which characterise the
X-ray sources and which are delivered by our GCA pipeline: the 
spectral hardness ratio and the extent of the source.
The hardness ratio, $HR$, is defined by
$HR = {H - S \over H + S}$
where $H$ is the hard and $S$ the soft band source count rate.
Both rates are determined using the same aperture radius. The bands 
are defined on the basis of the {\sf ROSAT} PSPC energy channels:
from 10 to 40 (for the soft band) and from 52 to 201 (for the hard
band). These intervals correspond roughly to the energy
bands 0.1 - 0.4 keV and 0.5 to 2.0 keV, respectively. For galaxy clusters
we can calculate the expected hardness ratios from the spectral
properties for given temperature, redshift, and interstellar hydrogen 
column density, $n_H$. For clusters with temperatures above 2 keV 
the hardness ratio depends mostly on the interstellar column density. 
Using the observed $n_H$ from Dickey and Lockman (1990) we can
determine the expected hardness ratio approximately 
(assuming a temperature of 5 keV and a metallicity of 0.3
solar) and apply this parameter in our cluster identification process.

\begin{figure}[ht!]
\figurenum{4}
\plotone{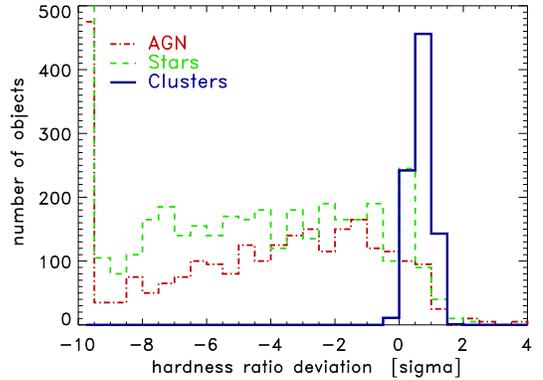}
\caption{Hardness ratio parameter distribution of various classes 
of X-ray sources: galaxy clusters, AGN, and stars.  
The X-axis shows the deviation of the observed hardness ratio from 
the expectation for a galaxy cluster given a temperature of 5 keV and
the interstellar absorption in the line-of-sight as taken from
Dickey \& Lockman (1990). The unit of the deviation is given in 
$\sigma$-values of the hardness ratio uncertainty. 
A substantial fraction of the non-cluster sources
can be distinguished by their hardness ratio from clusters.\label{fig4}}
\end{figure}

We make use of the hardness ratio in the following way. We compare 
the expected HR values to the observed ones and
their estimated uncertainties. We illustrate the usefulness of this parameter
in Fig.~\ref{fig4}, where we compare the hardness 
ratio deviations for different types of X-ray sources.\footnote
{The stellar and AGN X-ray sources used here have been compiled during
our source identification process from X-ray sources above
the flux limit with a safe identification as stars or AGN from
the literature.} Galaxy clusters
populate the high end of the hardness ratio distribution.
The stars are mostly softer than the clusters and the majority of
the AGN sources can be distinguished from clusters as been too soft.
We conservatively mark an 
X-ray source to be too soft to originate from a cluster, when the observed value is
more than $3\sigma$ away from the expectation. 

Compared to the majority of the sources which are mostly stars
and AGN, cluster sources are extended. Therefore the finding of 
a source extent is a strong indication for a cluster candidate. 
We cannot rule out a cluster nature of a source, however, if we do not find
an extent, since not all X-ray cluster sources at higher redshift can be
spatially resolved in the {\sf ROSAT} Survey. We search for a source
extent by applying a Kolmogorov-Smirnov (KS) test to estimate the probability 
that the photon distribution of the source is consistent with a point
source. A probability less than 0.01 is taken as sign of a significant
source extent. KS tests applied to X-ray sources with an identification
as stars or AGN return a false classification rate of not being consistent with
a point source with a probability less than 0.01 on the 5\% level. 

\begin{figure*}[ht!]
\figurenum{5}
\begin{centering}
\includegraphics[width=1.0\textwidth]{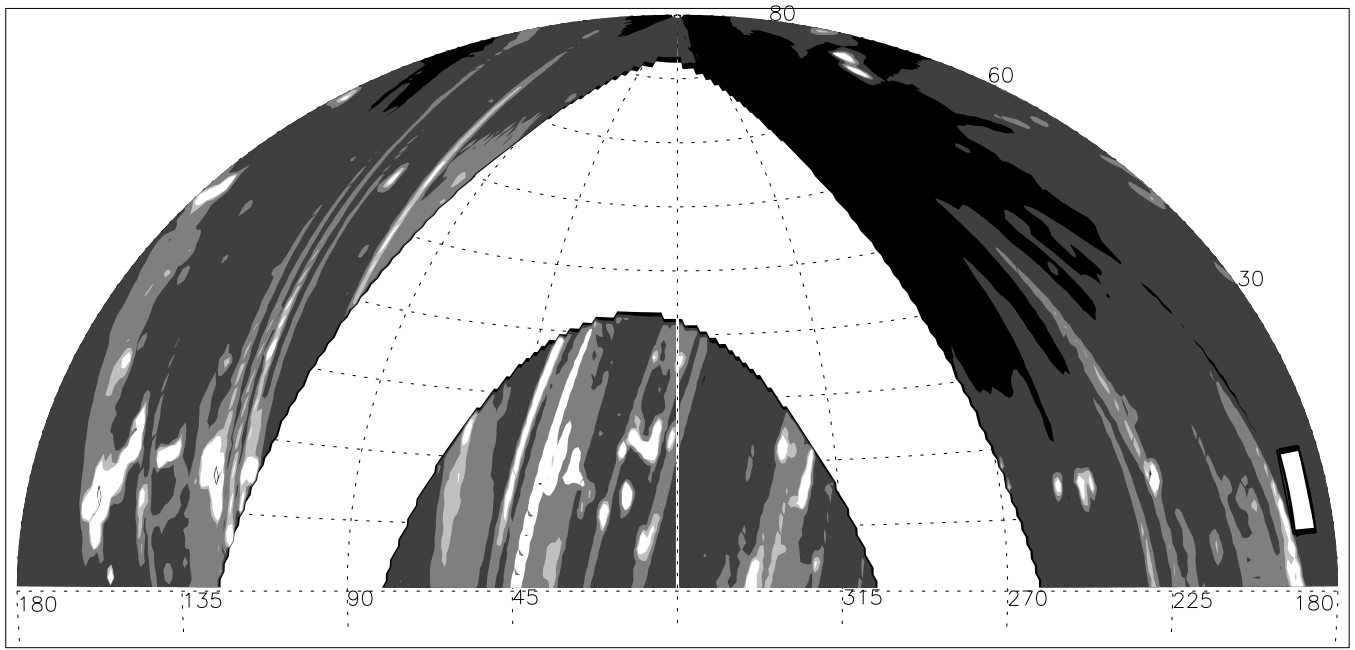}
\end{centering}
\caption{Sensitivity map of RASS III for the region of the 
{\sf NORAS II} survey in an equatorial coordinate system (J2000).
We show the number of photons detected for a source at the
nominal flux limit using five levels of increasing grey scale
for photon numbers $ < 15$ , $15 - 20$ , $20 - 30$, $30 - 60$, and $> 60$,
respectively.  The white box shows the Virgo cluster region
(RA: 185 - 191.25$^o$, DEC: 6 - 15$^o$). The data of the map are made
available on-line as explained in Appendix B.\label{fig5}}
\end{figure*}

The likely cluster candidates were selected from the flux 
limited list of 3902 X-ray sources based on our experience gained 
with {\sf REFLEX }, where identical selection criteria had been
used. First of all, we cannot rely 
on the detection of extended X-ray emission for the selection, 
since not all clusters are resolved in the survey with the 
survey PSF. For this reason we considered all X-ray sources
and took the following information into account:
the X-ray properties of the sources determined from
the RASS, firm object identifications listed in 
NED\footnote{see http://ned.ipac.caltech.edu/} at the 
source positions, digital sky images
obtained from DSS\footnote{see http://archive.stsci.edu/dss/},
and any relevant information from the literature. An inspection
of DSS images with superposed contour images of the
X-ray surface brightness turned out to be very helpful. 
For X-ray sources already clearly identified in the literature 
we adopt this classification if we have no doubt about it after an 
inspection of the observational information.  

In B\"ohringer et al. (2013), which described the construction of the 
{\sf REFLEX II} survey, we provided a detailed description of the
selection scheme including an overview by means of a flow chart of
the selection process (see their Fig. 9). To discard an X-ray
source from the cluster candidate list in a first step, we typically
use at least two negative criteria: (i) the source is too soft and is consistent with
being a point source, (ii) the source is a point source and coincident with
a bright star, known galactic X-ray source, or AGN, (iii) the source is flagged as 
extended and coincident with a nearby galaxy, galactic HII region, or supernova
remnant, (iv) the source is flagged as extended but best explained by multiple sources
with a hardness ratio or a morphology inconsistent with being a cluster. 
For some point sources with spectral properties not inconsistent 
with a cluster, but no trace of a galaxy concentration on the optical images,
we are left with no classification. In case there is any weak indication of
some faint galaxies at the X-ray center or for completely blank fields
we have taken deeper images as part of our observation runs. 
Inspecting more than 40 such borderline cases that had not been flagged as 
promising candidates we did not find a cluster. Since most promising fields 
have been targeted, we are convinced
that we have reached a high completeness in our cluster identification of
the flux limited source list. 
On the positive selection side we could be more generous to include weak
cluster candidates since in the follow-up spectroscopic identification, 
which is described below, the false clusters are revealed. 

In some cases we found that the cluster X-ray emission was contaminated 
by X-rays from point sources. In this case we did the best effort to deblend 
the point source from the cluster emission and added the estimated uncertainty 
of the deblending procedure to the flux error. Before deciding to
deblend, it is important to distinguish between truly contaminating point
sources and cluster substructure, where the latter is considered part of
the genuine cluster emission. We therefore deblended only sources which 
can be recognised as non-cluster emission by having a clearly separable
point source nature, by showing an obvious counterpart for the X-ray source,
or showing a distinctly different local hardness ratio. A nice example for
the latter case is shown in Fig. 3 in Chon \& B\"ohringer (2012).  
In another set of sources inspected during the source identification process,
there was either a previously identified group or cluster of galaxies
near the X-ray position or a cluster was well visible in the digital sky
images, but it was obvious that the X-ray emission was likely dominated
by an AGN, while the possible cluster emission fell below the flux 
threshold of the survey. In these cases we removed the sources from the catalogue,
and plan to give a list of them with the publication of the cluster catalogue,
as done for {\sf REFLEX I} (B\"ohringer et al. 2004).

The confirmation of the cluster's nature and redshift measurement 
by follow-up spectroscopic observations were described in Guzzo et al. 
(2009) for {\sf REFLEX I} and in Chon and B\"ohringer (2012) for {\sf REFLEX II}
and they apply in a similar way to the NORAS project. For the {\sf NORAS II} 
survey, observations were performed mainly at the German-Spanish Calar Alto Observatory
at the 3.5 and 2.2m telescopes. For {\sf NORAS I} we also conducted
observations at Mt. Hopkins (B\"ohringer et al. 2000). Whenever possible, 
we  used multi-slit spectroscopy, which typically provided  
about 7 cluster galaxy redshifts. To make the observing runs most efficient,
in particular at the smaller telescopes,
we also used long slit observations with two or three targets per slit.
For these cases it was important to include the BCG as one of the targets,
which helped very much to get a unique cluster redshift. The typical galaxy redshift
uncertainty is 50 - 60 km/s for our follow-up observations.

From the selection and identification of the X-ray source list we obtained
a catalogue of currently 853 galaxy clusters, with 7 additional clusters 
in the Virgo region. For this sample we have not yet applied a lower limit
in the detected number of source photons of 20. 21 clusters in the sample
have less than 20 and 7 less than 15 source photons.  

\section{Survey selection function}

To provide a model of the survey for the further data analysis
the survey selection function has to be known. To characterise 
the survey sensitivity of {\sf NORAS II} for each position in the sky,
we calculate the flux limit per photon based on  
the exposure time of RASS III and the interstellar column density taken 
from Dickey \& Lockmann (1990). For the count rate to flux conversion 
we use the same conversion calculation as used in section 2 for 
the determination of the nominal source fluxes, with the same assumptions 
(ICM temperature of 5 keV and metallicity of 0.3 solar).
We call this the sensitivity map of the survey, 
and show it in Fig.~\ref{fig5}. The map shows the number of detectable 
photons for sources at the nominal flux limit in pixels with a resolution
of one square degree. There are stripes of low sensitivity, which
are due to low exposure regions suffering from shut down times of 
the {\sf ROSAT} detectors during the crossing of the 
radiation belts. {The {\sf NORAS} map has
fewer of these low exposure areas} than the corresponding southern sky map.
This results in a higher average exposure of the {\sf NORAS} region
compared to the {\sf REFLEX} survey.  The deepest exposure is at 
the North Ecliptic Pole region at 18$^h$ and +66$^o$ 33' 38.5''.  

\begin{figure}[ht!]
\figurenum{6}
\plotone{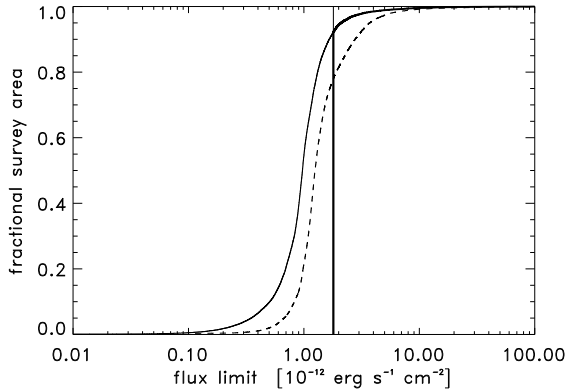}
\caption{Effective sky coverage
of the {\sf NORAS II} sample. The thick line gives the effective sky area
for the nominal flux limit of $1.8 \times 10^{-12}$ erg s$^{-1}$ cm$^{-2}$
and a minimum number of 20 photons per source. The solid line without
the vertical cut at a flux of $1.8 \times 10^{-12}$ erg s$^{-1}$ cm$^{_2}$
shows the 20 photon limit. The dashed line shows the equivalent 
20 photon limit for the {\sf REFLEX II} Survey.\label{fig6}}
\end{figure}

In the next step we calculate the nominal flux limit for the 
survey as a function of sky position imposing a limit for
the minimum number of source counts for a robust detection.
Different values for this limit could be chosen depending on
the requirements for the quality of the source properties. 
We have chosen a minimum limit of 20 source 
photons for several reasons.
Decreasing this photon number, the flux and hardness 
ratio uncertainties increase relatively fast.
Also the region of the survey, where the nominal flux limit 
is not reached, increases. While these uncertainties can in 
principle be taken care of, accurate estimates of the uncertainties
become increasingly difficult below a photon number of 20. On the 
other hand few sources are lost by imposing this cut, as 
described above. Therefore we lose in general more than we gain
by lowering this limit. The cumulative sky area covered by the 
survey as a function of the nominal flux limit for a minimum detection
of 20 photons is shown in Fig.~\ref{fig6}. We also provide
the corresponding curve for {\sf REFLEX II}. We note that over
92.2\% of the sky area the nominal flux limit of $1.8 \times 10^{-12}$
erg s$^{-1}$ cm$^{-2}$ is reached. Only the smaller 
remaining part has a higher limiting flux. The figure also
shows that in the northern sky the area in which the nominal 
flux-limit is reached is larger compared to that of the 
{\sf REFLEX} survey. We provide the data of the curve
in numerical format in the on-line material as
described in the Appendix.         

\begin{figure}[ht!]
\figurenum{7}
\plotone{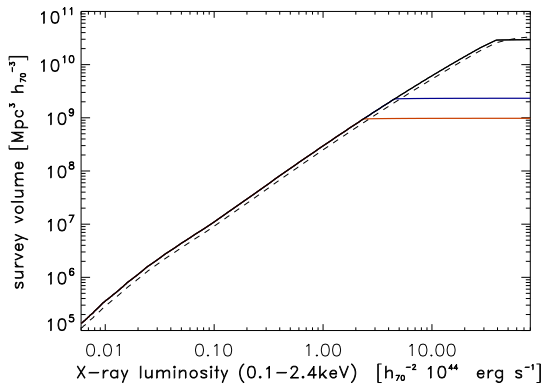}
\caption{Effective survey volume  as a function of X-ray
luminosity of the {\sf NORAS II} survey. The solid curve shows the  
survey volume for a redshift limit of $z = 0.8$ (upper curve), a limit of 
$z = 0.3$ (middle curve) and a limit of $z= 0.22$ (lower curve), respectively.
The dashed line shows the {\sf REFLEX II} survey volume for a redshift
limit of $z = 0.8$.\label{fig7}}
\end{figure}

\begin{figure}[ht!]
\figurenum{8}
\plotone{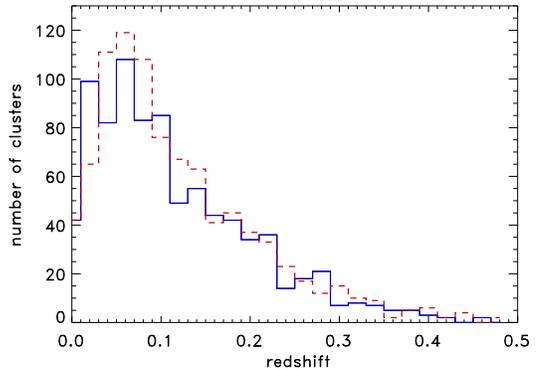}
\caption{Redshift distribution of the {\sf NORAS II} clusters 
(solid line). This function is compared to that of the {\sf REFLEX II} sample 
(dashed line).\label{fig8}}
\end{figure}

The construction of the X-ray luminosity function and several 
cosmological studies require the knowledge of the 
survey selection statistics as a function of cluster luminosity. 
We provide this information in the 
form of the minimum cluster luminosity required for a detection 
as a function of sky position and redshift. We will refer to this 
multi-dimensional function as the selection mask of the survey. This mask
provides for example the direct recipe to select clusters from a cosmological N-body 
simulation to create a mock survey sample. The mask was applied 
in exactly this way in our study of the {\sf REFLEX II} power spectrum 
in Balaguera-Antolinez et al. (2011). 

To determine the selection mask we calculate for each sky pixel 
the limiting luminosity 
as a function of redshift. While this calculation
starts with a 5 keV ICM spectrum for the count rate - flux conversion,
we iteratively include the estimated temperature, the ``K-correction''
and the missing aperture flux in the calculation.
The mask is tabulated for 160 redshifts out to $z = 0.8$ and for
21666 sky pixels. The data for the mask are given in the on-line
material in a form further described in the Appendix. 

Another useful statistic, required for example to determine the 
cluster X-ray luminosity function, is the effective survey volume 
probed by {\sf NORAS II} as a function of X-ray luminosity. We show
this function in Fig.~\ref{fig7} for the fiducial $\Lambda$CDM cosmological
model defined in the introduction. We limited the volume calculations 
to redshifts up to $z = 0.8$ (shown as solid line) and alternatively to
redshift limits of $z = 0.3$ and $z = 0.22$  (shown by shown
by dashed lines), respectively. For a luminosity of $L_X = 5 \times 10^{44}$
erg s$^{-1}$, which is not far from $L^{\star}$ of the X-ray luminosity function,
we reach a survey volume of $\sim 2.5$ Gpc$^3$. The redshift limit of $z = 0.3$
approximately corresponds to this limiting luminosity. 

\section{Statistical properties of the cluster sample}

Fig.~\ref{fig8} shows the redshift distribution of the {\sf NORAS II}
clusters in comparison to those of {\sf REFLEX II}. The redshift
distributions are quite similar. The median redshift for the {\sf NORAS II}
sample is $z = 0.102$ and the mean $z=0.125$. There is a long tail of 
distant clusters out to redshifts of $\sim 0.5$.

\begin{figure}[ht!]
\figurenum{9}
\plotone{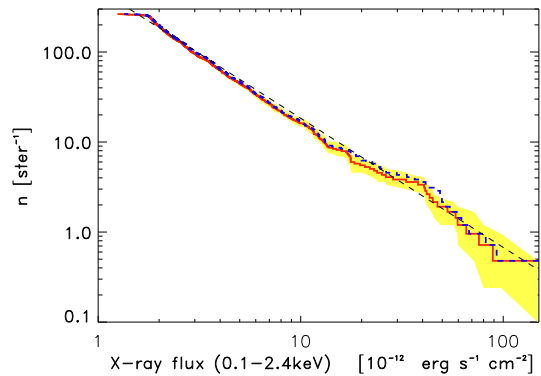}
\caption{LogN-logS distribution of the clusters in the {\sf NORAS II} survey.
The solid line shows the function for measured fluxes and the dashed
line for those of the fluxes inside $r_{500}$, corrected for missing flux.
The dashed, straight line shows a fit to a power law with a logN-logS slope
of -1.44. The shaded region shows the 1$\sigma$ uncertainties obtained from
bootstrap simulations.\label{fig9}}
\end{figure}

The cluster number counts per unit sky area as a function of limiting flux,
the so-called logN-logS distribution, for {\sf NORAS II}
is shown in Fig.~\ref{fig9}. It is restricted to clusters with more
than 20 photon counts.
The source density is calculated with a sky area normalisation 
derived from the nominal flux, $F_n$, and the sensitivity map shown in 
Fig.~\ref{fig5}. The flux values used in the curve are 
the corrected observed flux, $F_X$ (solid line), and the 
flux corrected for missing flux inside $r_{500}$ (dashed line). The best
fitting slope for the observed LogN-logS function is about $-1.44$, slightly 
steeper than the fit to the logN-logS function of {\sf REFLEX II}.

To estimate the statistical uncertainties of the LogN-logS function
we performed bootstrap simulations by resampling from the observed cluster
flux distribution, allowing the total number of clusters
in the sample vary within its Poisson errors. We obtained 1$\sigma$ limits 
for the uncertainties by determining the 16\% and 84\% percentiles of the
cluster number densities as a function of flux from 1000 simulations
as shown in Fig. 9.

\begin{figure}[ht!]
\figurenum{10}
\plotone{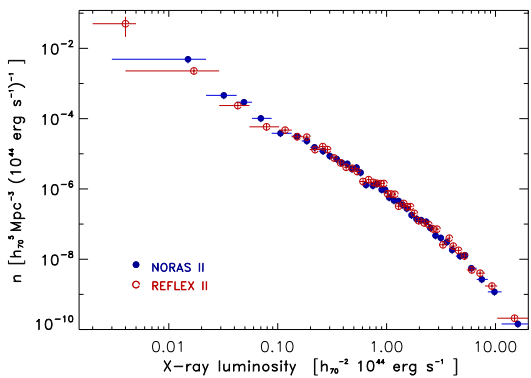}
\caption{X-ray luminosity function for the clusters from the {\sf NORAS II} 
survey (filled symbols). For comparison we show also the results from
{\sf REFLEX II} (open symbols).\label{fig10}}
\end{figure}

\section{The X-ray luminosity function}

The most important census of the population of X-ray {\bf selected} clusters
in the Universe is the X-ray luminosity function (XLF).
This function is determined from the 
catalogue of clusters and the survey selection function in the 
form of the effective survey volume as a function of X-ray luminosity 
as shown in Fig.~\ref{fig7} . We use a source detection count cut of 
minimum 20 photons for the selection. Then the binned differential
X-ray luminosity function is given by

\begin{equation}
{dn(L_X) \over dL_X} = {1 \over \Delta L_X} \sum_i {1 \over V_{max}(L_{X_i})}
\end{equation}

where $V_{max}$ is the effective detection volume, $\Delta L_X$ is 
the width of the luminosity bin, and the sum includes
all clusters in the bin. Fig.~\ref{fig10} shows 
the XLF derived for a binning with 20 
clusters per bin, except for the bin at the lowest X-ray luminosity
which contains only 18 clusters.
The luminosity values used in the construction of the XLF
are the luminosities inside $r_{500}$, corrected for missing flux
in the 0.1 - 2.4 keV rest frame band. The sub-sample used 
for the analysis covers the redshift 
range $z = 0$ to $0.4$.
The error bars for the XLF given in the figure are the Poisson
uncertainties for the number of clusters per bin.
The XLF covers more than three orders of magnitude in luminosity
and comprises objects from small groups to the largest clusters
we know. 

In the figure we also compare the {\sf NORAS II} XLF to 
that of {\sf REFLEX II} and find excellent agreement within
the Poisson uncertainties. We do not expect cosmic variance to
play a significant role here, except for the lowest luminosity bin,
as shown in B\"ohringer et al. (2002). The redshift range used
for the construction of the XLF is rather large, $z = 0$ to $0.4$,
and thus one may wonder about the effects of redshift evolution.
For the case of {\sf REFLEX II} we have looked into the evolution
of the XLF in detail and found no significant effect in this
redshift range (B\"ohringer et al. 2014). We can attribute
the lack of observed evolution to the fact that an expected 
mild evolution in the 
mass function is compensated by an evolution of the 
mass - X-ray luminosity relation.

\section{Constraining cosmological parameters}

The most immediate application of these results to cosmology
is the use of the cluster abundance encoded in the XLF
to constrain cosmological parameters by a comparison of the
observational data with theoretical predictions for various
cosmological models. We apply here to the {\sf NORAS II} 
data set exactly the same approach that we used in
B\"ohringer et al. (2014) for the {\sf REFLEX II} sample.
The only difference is the calculation of the
large-scale structure matter power spectrum by means
of the program {\sf CAMB} (Lewis et al. 2000)\footnote{CAMB
is publicly available from http://www.camb.info/CAMBsubmit.html} 
instead of the power spectrum
by Eisenstein \& Hu (1998) used previously. This change makes
a difference in the results of less than one percent.
In brief the modelling of the cluster luminosity function
involves the following steps. We calculate theoretical 
predictions for the 
cluster mass function from structure formation theory including
the mass function formulas by Tinker et al. (2008). We use
the following X-ray luminosity - mass relation (also used
in B\"ohringer et al. 2014)

\begin{equation}
L_{500}{\rm (0.1 - 2.4 ~keV)} = 0.1175~ M_{200}^{\alpha_{sl}}~ 
h^{\alpha_{sl}-2} ~~~E(z)^{\alpha_{sl}} ~~~~~~,
\end{equation} 

to convert the mass function to a cluster XLF
taking account of the uncertainties and the scatter of
the relation. This relation is based on observations 
and chosen such that it is consistent with our 
results from the REXCESS study in Pratt et al. (2009),
with the work of Vikhlinin et al. (2009) and with our 
earlier results reported in Reiprich \& B\"ohringer (2002).
The resulting, predicted XLF is compared to the observations
of {\sf NORAS II} by means of a maximum likelihood method.
We allow for measurement uncertainties for the X-ray fluxes and
luminosities of 20\%.

\begin{deluxetable}{l|lr}
\tablecaption{Default cosmological model parameters.}
\tablehead{
\colhead{Parameter} & \colhead{Explanation} & \colhead{Value} 
}
\colnumbers
\startdata
$h_{100}$ & {\rm Hubble~ parameter}  & 0.70 \\
$\Omega_b$ & {\rm baryon~ density} &  0.045 \\
$n_s$ & {\rm Primodial~ P(k)~ slope} & 0.96 \\
$\alpha_{sl}$ & $L_X - M$~ {\rm relation~ slope}& $1.51 \pm 7\% (1\sigma$)\\
$n_0$ & $L_X - M$~ {\rm relation~ norm.}  & $0.1175 \pm 14\% (1\sigma$)  \\
 {\rm mass~bias} & {\rm X-ray~mass~underestimation} & 0.1 \\
\enddata
\tablecomments{We use a flat $\Lambda$CDM cosmological model 
with the parameters given in the Table.}
\end{deluxetable}

\begin{figure}[ht!]
\figurenum{11}
\plotone{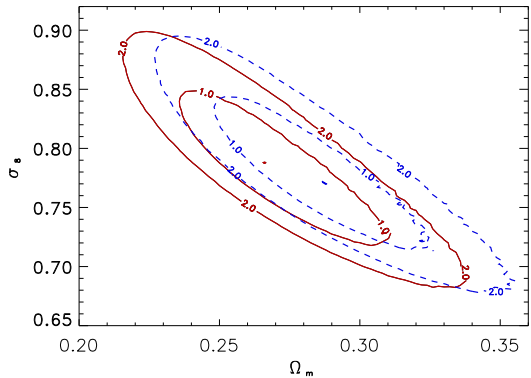}
\caption{Constraints on the cosmological parameters $\Omega_m$ and
$\sigma_8$ obtained from the {\sf NORAS II} cluster sample (solid 
lines, 1 and 2~$\sigma$ contours) compared to the results from the
{\sf REFLEX II} survey (dashed lines).\label{fig11}}
\end{figure}

We concentrate here on the constraints of the cosmological 
parameters $\Omega_m$  and $\sigma_8$, to which the 
galaxy cluster XLF is most sensitive. The other cosmological
parameters were fixed to the same values as in our cosmological
analysis of {\sf REFLEX II} (B\"ohringer et al. 2014) 
with negligible neutrino mass.
The cosmological and scaling relation parameters we use are 
summarised in Table 1.

As in B\"ohringer et al. (2014) we use a generous marginalisation
over the uncertainties in the scaling relation given in Eq. 1,
with errors on the $L_X - M$ relation slope of $\pm 7\%$, and for the
normalisation of $14\%$, taken as 1$\sigma$ uncertainties.
The uncertainty in the normalisation is equivalent to a
bias in the mass calibration for the relation. We allow
for a scatter in the relation of 30\%. In addition
we also assume that the mass calibration, 
which is based on hydrostatic mass determinations using X-ray data,
is biased low by 10\%. For a detailed discussion of the
effect of these uncertainties and the justification for 
their values see B\"ohringer et al. (2014).

Fig.~\ref{fig11} shows the constraints we obtain for the 
two cosmological parameters $\Omega_m$ and $\sigma_8$.
We compare the  {\sf NORAS II} results in the figure to 
those of {\sf REFLEX II} and we find excellent agreement 
within the statistical uncertainties. We will not discuss the
cosmological implications of these results here, since this
can be found in our previous publications B\"ohringer et al. 
(2014, 2015) and B\"ohringer \& Chon (2016).

\section{Discussion}

\begin{figure}[ht!]
\figurenum{12}
\plotone{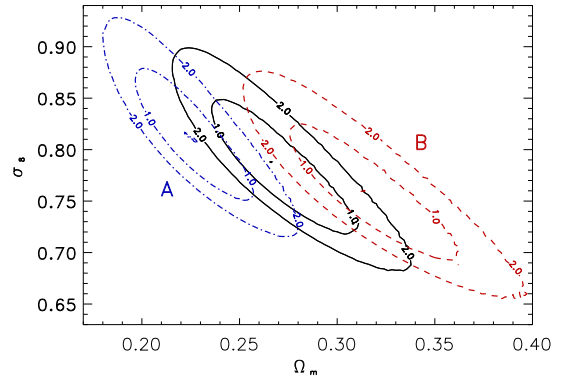}
\caption{Effect of the sample completeness on the cosmological 
parameter constraints. We show the variation of the marginalised 
constraints on $\Omega_m$ and $\sigma_8$ for a sample depleted
to 80\% (A), a nominal sample (100\%, solid lines) and an artificially enriched
sample (130\%, B) of the {\sf NORAS II} survey.\label{fig12}}
\end{figure}

Most of the systematics that affect these studies has been discussed
in B\"ohringer et al. (2014). One aspect we investigate here
in more detail than what has been done previously 
is the question how the results are affected by
an incompleteness of the cluster sample, since we are still missing
3\% of the redshifts for the {\sf NORAS II} clusters, which are therefore
not included in the cosmological analysis.
Thus we have repeated the above exercise of cosmological 
parameter constraints with two cluster samples, where 
the cluster catalogue for sample A has been reduced to 80\% of the
original catalogue (by excluding the last 20\% of the clusters
in right ascension order). For sample B the first 30\% of the clusters in
right ascension order have been counted twice. The parameter
constraints for the original sample and for the two manipulated
samples are shown in Fig.~\ref{fig12}. The changes of the results shown
are much larger than what we expect for any realistic incompleteness,
to illustrate the effect more clearly.

The dependence of the results on the sample size can be roughly 
parameterised by ${d log \Omega_m \over dlog S} \sim 0.7$ and  
${d log \sigma_8 \over dlog S} \sim -0.15$. This is only valid close 
to the nominal value as the dependence is not linear.  
For {\sf NORAS II}, where 3\% of the redshifts are still missing, 
this would imply an increase of $\Omega_m$ and $\sigma_8$ 
for the full sample of about 2\% and -0.5\%, respectively.
Overall we expect that the sample could be incomplete in the cluster
detection by up to about 6\%, which would introduce a bias of
about 4\% and  1\%, respectively, which is still well within the
statistical uncertainties. 

This test was performed under the assumption that the incompleteness
is statistically random. However, the clusters remaining without redshifts 
so far may have highly biased properties - at least most of them are close
to the flux limit and not nearby. In the Appendix we therefore
also explore the effect when the missing clusters have preferentially
low or high X-ray luminosities. The results shows that the incompleteness 
effect shifts the results in these two cases in opposite directions.
Therefore it is dangerous to make any correction for the incompleteness
if we do not know the properties of the clusters with missing redshifts.
For this reason we have not used the results shown in Fig. 11 to 
apply a correction to our cosmological results.

\section{Summary and Conclusion}

One of the major objectives in compiling the {\sf NORAS II} cluster sample
was to make the northern RASS cluster sample compatible with {\sf REFLEX II}.
Therefore we took care to construct the sample with the same selection
parameters and aimed for the same quality as for the {\sf REFLEX II} Survey.
The good agreement of all the results on the survey statistics and 
the cosmological parameters between the two surveys shows that we
have reached this goal. Therefore we can now combine the samples
for the two hemispheres to build a homogeneous
all-sky cluster sample from the ROSAT Survey outside the zone-of-avoidance. 
This is the largest and best defined sample of X-ray luminous clusters 
in the Universe. It comprises 1722 clusters in a sky area of $\sim 8.25$
ster. We have used this combined survey already recently to study the
statistics of rotation measures of background radio sources in the 
line-of-sight of the sample clusters to obtain information on the 
magnetic fields in galaxy clusters (B\"ohringer et al. 2016).

One of the most exciting applications of this sample is the study of
the large scale matter distribution of the nearby Universe out to redshifts
up to $z \sim 0.3$. In B\"ohringer et al. (2015) we have already studied
the mass distribution in the nearby Universe in the southern sky. 
With the completion of the all-sky survey we can study the very interesting 
aspect of the dipole in the density distribution of the Universe
that gives rise to the local bulk flow. Having reached a homogenisation
of the ROSAT cluster sample across the sky, this data set is now perfectly
suited for this study.

\acknowledgments
We like to thank the {\sf ROSAT} team at MPE for the support with the data
of the RASS and the staff at the German-Spanish Calar Alto observatory
for the technical support during the observing runs. 
H.B. and G.C. acknowledge support from the DFG Transregio Program TR33
and the Munich Excellence Cluster ''Structure and Evolution of the Universe''.  
G.C. acknowledges support from Deutsches Zentrum f\"ur Luft- und Raumfahrt
under grant 50OR1601. This research made use of the NASA/IPAC Extragalactic 
Database (NED), which is operated by the Jet Propulsion Laboratory 
under contract by NASA. We like to thank the anonymous referee for helpful
comments.







\appendix

\section{Effect of sample incompleteness}

\begin{figure}[ht!]
\figurenum{A1}
\plotone{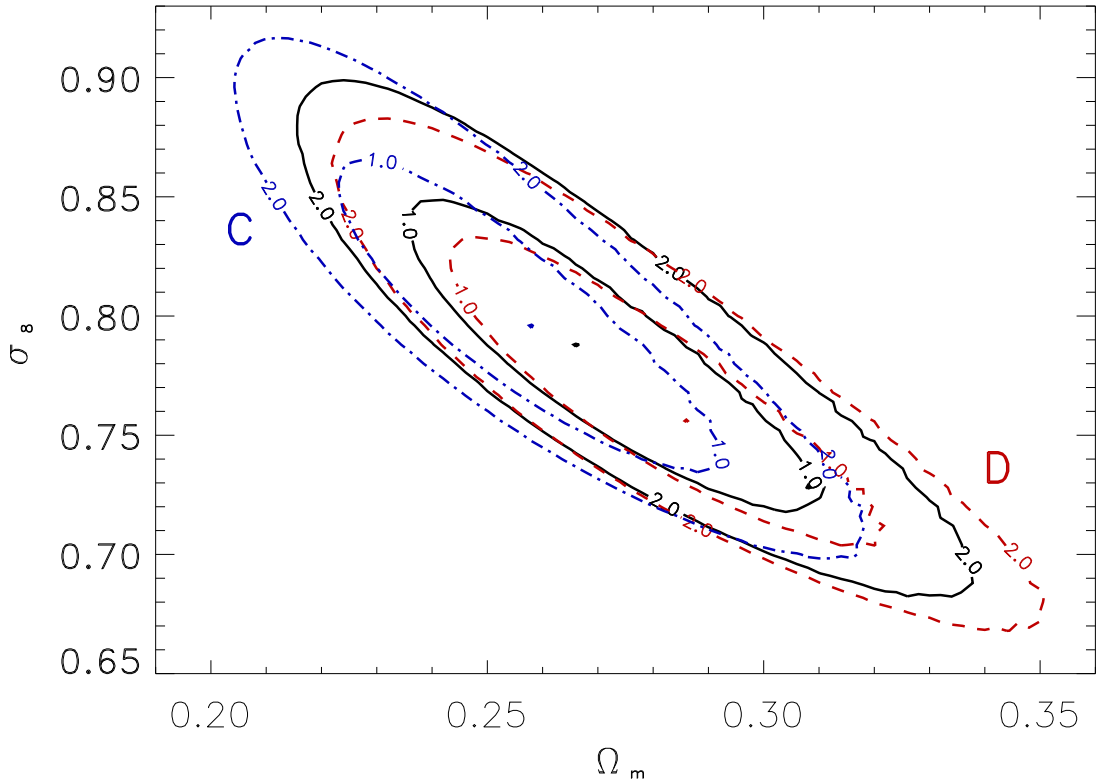}
\caption{Effect of an incomplete X-ray cluster sample on the derived
cosmological parameter constraints. For sample C we have removed 3\%
low luminosity clusters, while sample D is depleted by clusters
at high X-ray luminosity.\label{fig13}}
\end{figure}

In section 8 we discussed the effect of sample incompleteness
on the derived cosmological parameter constraints. In these
studies we have removed clusters from or added clusters to the
sample in a statistically random fashion. In reality we expect 
the last 3\% of the clusters left without redshift not to have
average properties, but rather to be for example mostly close
to the flux limit, not very nearby, not the easiest to recognise
in optical images. Therefore it is interesting to explore the 
incompleteness effect on the cosmological constraints 
when the missing clusters have preferentially low or high 
X-ray luminosity. To this end we statistically removed 3\% of 
the clusters in the luminosity range $L_X = 0.25 - 1 \times
10^{44}$ erg s$^{-1}$ for sample C. Note that clusters with 
$L_X < 0.25 \times10^{44}$ erg s$^{-1}$ are not used for
the comsmological parameter constraints, because the
$L_X -M$ relation is not well calibrated in this range. 
For sample D we removed  3\% of 
the clusters in the luminosity range $L_X \ge 3 \times10^{44}$ erg s$^{-1}$.

The results are shown in Fig.~\ref{fig13}. The best fitting parameters move
in opposite directions in the $\Omega_m$ - $\sigma_8$ parameter
plane. This can be understood from an inspection of Fig. B.1
in B\"ohringer et al. (2014). Removing high luminosity clusters
makes the luminosity function steeper, which is better fit by a lower
$\sigma_8$, while $\Omega_m$ will be increased to compensate 
for the lowering of the normalisation from a decrease of $\sigma_8$.
The opposite effect is observed by removing low luminosity clusters
which is flattening the luminosity function.

\section{Survey Sensitivity and Selection Function}

In this section we provide the numerical data associated to
Figs. 5 and 6 and the data of the survey mask. Table B2 
provides the information given in Fig. \ref{fig5},
the equatorial and galactic coordinates of the sky pixels 
for the epoch J2000 and the number of photons that can be detected in
the RASS at that pixel for a source at the limiting flux, $F_X = 1.8 \times 10^{-12}$
erg s$^{-1}$ cm$^{-2}$.

\begin{deluxetable}{rrrrrr}
\tablecaption{Sensitivity map of the NORAS II Survey.}
\tablehead{
\colhead{RA} & \colhead{DEC} & \colhead{S} & \colhead{size} 
& \colhead{g-long}  & \colhead{g-lat}} 
\colnumbers
\startdata
  0.5000 & 89.5000 & 54.217 & 0.008726 & 122.8131 & 26.6396 \\
  1.5000 & 89.5000 & 55.215 & 0.008726 & 122.8226 & 26.6378 \\
  2.5000 & 89.5000 & 55.215 & 0.008726 & 122.8322 & 26.6362 \\
  3.5000 & 89.5000 & 55.215 & 0.008726 & 122.8418 & 26.6347 \\
  4.5000 & 89.5000 & 55.046 & 0.008726 & 122.8515 & 26.6334 \\
\enddata
\tablecomments{Columns (1) and (2) provide the equatorial coordinates of the
sky pixel for J2000 and columns (5) and (6) the corresponding galactic coordinates.
Column (3) gives the number of photons that can be detected in that sky pixel
for a source at the nominal flux limit. Column (4) gives the size of the
sky pixel in square degrees. We show the first five lines of the Table, the
full table is provided with the on-line data.}
\end{deluxetable}

Table B3 provides the data on the sky coverage as a function of flux
shown in Fig. \ref{fig6}. It gives the fraction of the sky covered for
a flux limit as quoted in the first column, based on a detection of
at least 20 photons.

\begin{deluxetable}{rr}
\tablecaption{Survey mask for NORAS II.}
\tablehead{
\colhead{flux} & \colhead{counts} \\}
\colnumbers
\startdata
     0.010 & 0.001186 \\
     0.012 & 0.001215 \\
     0.014 & 0.001245 \\
     0.014 & 0.001274 \\
     0.020 & 0.001303 \\
\enddata
\tablecomments{Column (2) gives the fraction of the survey
area of {\sf NORAS II} (13519 deg$^2$) covered to the limiting flux quoted
in column (1) in units of $10^{-12}$ erg s${-1}$ cm$^{-2}$
in the 0.1 to 2.4 keV band for a detection of at least 20
source photons.}
\end{deluxetable}

Table B4 provides the the numerical values of the survey mask. It lists for
each sky pixel the minimum X-ray luminosity required for a detection in the
RASS with 20 source photons for 160 redshifts from $z = 0.005$ to $z = 0.8$.

\begin{deluxetable}{rrrrrr}
\tablecaption{Survey mask for NORAS II.}
\tablehead{
\colhead{RA} & \colhead{DEC} & \colhead{z = 0.005} & 
\colhead{z = 0.01} & ... & \colhead{z = 0.8} }
\colnumbers
\startdata
0.5000 & 89.5000 & 0.00873 & 0.00145 & ... & 38.90970 \\
1.5000 & 89.5000 & 0.00873 & 0.00145 & ... & 38.90970 \\
\enddata
\tablecomments{The Table gives for every sky position with
RA and DEC given in columns (1) and (2) the minimim
X-ray luminosity (0.1 - 2.4 keV) to be detected with 20 source photons
at the redshifts given in the header. Here we give the
results only for 3 redshifts and show only the first two 
lines of the Table. The full Table, which is provided with the 
on-line data, lists the minimum luminosities for 160
redshifts between z = 0.005 and 0.8.}
\end{deluxetable}

\listofchanges

\end{document}